# Experimental study of the incoherent spectral weight in the photoemission spectra of the misfit cobaltate $[Bi_2Ba_2O_4][CoO_2]_2$

A. Nicolaou[1,2], V. Brouet[1], M. Zacchigna[3], I. Vobornik[3], A. Tejeda[2,4], A. Taleb-Ibrahimi[2], P. Le Fèvre[2], F. Bertran[2], S. Hébert[5], H. Muguerra[5], D. Grebille[5]

[1] Lab. Physique des Solides, Université Paris-Sud, UMR8502, Bât 510, 91405 Orsay (France)
[2] Synchrotron SOLEIL, L'Orme des Merisiers, Saint-Aubin-BP48, 91192 Gif-sur-Yvette (France)
[3] CNR - INFM, Lab. Nazionale TASC c/o Area Science Park, s.s. 14 Km. 163.5, I-34012 Basovizza (TS) (Italy)
[4] Institut Jean Lamour, CNRS-Nancy-Université-UPV-Metz, 54506 Vandoeuvre-les-Nancy (France)
[5] Laboratoire CRISMAT, UMR 6508 CNRS et Ensicaen, 14050 Caen (France)

Previous ARPES experiments in $Na_xCoO_2$ reported both a strongly renormalized bandwidth near the Fermi level and moderately renormalized Fermi velocities, leaving it unclear whether the correlations are weak or strong and how they could be quantified. We explain why this situation occurs and solve the problem by extracting clearly the coherent and incoherent parts of the band crossing the Fermi level. We show that one can use their relative weight to estimate self-consistently the quasiparticle weight Z, which turns out to be very small $Z=0.15\pm0.05$. We suggest this method could be a reliable way to study the evolution of correlations in cobaltates and for comparison with other strongly correlated systems.

In a Fermi-liquid, elementary excitations can be described as quasiparticle (QP), but with a weight Z that decreases as correlations increase, the remaining weight being transferred to incoherent excitations. Angle-resolved photoemission (ARPES) is a unique tool to observe both coherent and incoherent excitations. This often leads to a characteristic "peak-dip-hump" (PDH) structure of the spectra, where the peak corresponds to the QP and the hump (HP) to the incoherent excitations. The PDH is then a direct "image" of the correlation strength, but its interpretation is not always straightforward, as it can have quite different origins. In a strongly correlated metal, the QP band is typically renormalized by a factor $Z^{-1}$ and the HP corresponds to the residual lower Hubbard band [1]. Another type of PDH can be observed, even in a weakly correlated metal, due to the coupling between electrons and a collective mode of energy $\omega_0$, most frequently phonons [2]. It occurs in the vicinity of $\omega_0$ and, for moderate couplings, the dominant effect is a "kink" in the dispersion at $\omega_0$ [3]. The amplitude of the kink is directly related to the strength of the coupling, but $\omega_0$ is an independent energy scale.

In this paper, we describe an intermediate situation, where a PDH structure (see Fig. 1) is found with a dip at an energy $\omega_0=0.2eV$, larger than typical phonon frequencies, but smaller than the 1.2eV bandwidth predicted by LDA calculations [4]. This structure is observed in the misfit cobaltate $[Bi_2Ba_2O_4][CoO_2]_2$ (BiBaCoO, see [5] for details). A similar PDH is present in Na cobaltates [6,7], which contains identical $CoO_2$ slabs. The question arises on whether $\omega_0$ indicates the QP bandwidth or a "kink" energy. We will show that Z can change from 0.1 to 0.7, depending on this interpretation. Such an incertitude on Z is clearly inconclusive, which is highly regrettable, as correlations vary in an intriguing way in cobaltates with doping of the $CoO_2$ slabs [5,8,9] that would be interesting to document directly with ARPES. We solve this problem by using the redistribution of spectral weight between QP and HP as an indicator of the interaction strength. This establishes that the PDH corresponds to strong many-body effects, characterized by $Z=0.15\pm0.05$. Moreover, very similar lineshapes are observed in other important class of correlated metals, like manganites [10] or weakly doped cuprates [11]. This study therefore offers a new reference example and indicates methods to compare and classify these structures.

As the PDH of BiBaCoO occurs within the LDA band width, we first have to demonstrate that it is due to many-body effects and not to band structure effects. Indeed, there are 3 bands ($a_{1g}$ and 2 $e'_g$) from the Co $t_{2g}$ manifold in a 1eV window below $E_F$ and Qian et al. attributed the dip in the PDH they observe in $Na_xCoO_2$ to an anticrossing hybridization gap between $a_{1g}$ and $e'_g$ [7]. In the data presented here, we use the light polarization to select bands of different symmetry (polarization dependent spectra were also reported in [7]). Comparing the band structure in two high symmetry directions, we conclude that the PDH is the intrinsic many-body lineshape of the $a_{1g}$ band. Furthermore, we show that subtracting spectra with different light polarization is a *very efficient way to isolate the PDH structure*. Consequently, we are able to extract the dispersion, width and area of the QP and HP. We show that the small QP weight ($Z\approx 0.15$) rules out a simple coupling with phonons or another bosonic mode as the origin of the PDH, although it was often interpreted this way [6,7,12,13]. On the other hand, it naturally identifies $\omega_0=0.2eV$ as the QP band width, which is a clear

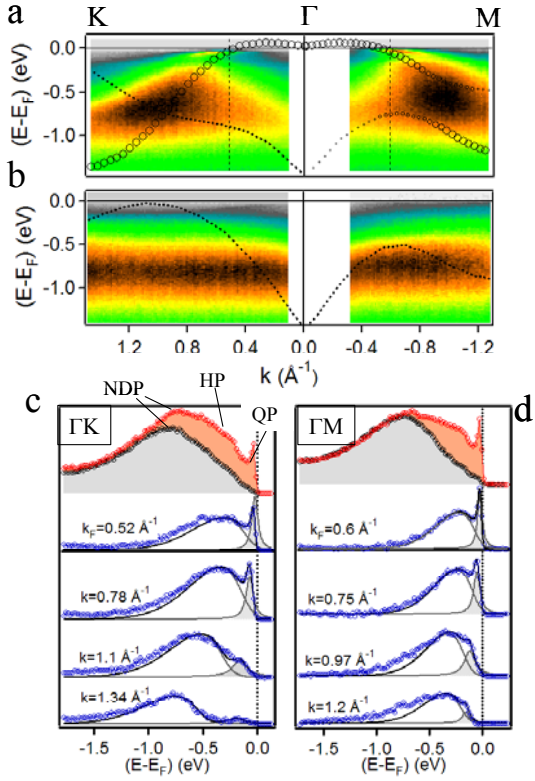

Fig. 1 : (a-d) Dispersion images along ΓK (left) and ΓM (right) with LH (a) and LV (b) polarizations. The LDA dispersion of even bands ($a_{1g}$ and $e'_{g2}$) are superimposed to the LH image and of the odd $e'_{g1}$ band to the LV image [4]. The $a_{1g}$ character is indicated by the size of the markers. (c-d) Top : Spectra at $k_F$ in LH (red) and LV (black) before subtraction along ΓK (left) and ΓK (right). Bottom (blue) : Difference spectra (LH-LV) at the indicated k values, fitted with a lorentzian cut by the Fermi function for the QP and an asymmetric function for the HP [16].

evidence for strong many-body effects in these systems.

The single crystals were prepared by a standard flux method and characterized by transport and magnetic measurements [5]. In Fig. 1, we show ARPES measurements taken at the APE beamline of ELETTRA [14], with a SCIENTA SES2002 analyser, an angular resolution of 0.2° and an energy resolution of ~20meV. The temperature was 20K, the photon energy 86eV and the beam was linearly polarized, either in the plane of incidence [Linear Horizontal (LH)] or perpendicularly [Linear Vertical (LV)] (the plane of incidence is defined by the incoming beam and the sample's surface normal). The sample was aligned by LEED and this alignment was confirmed by the periodicity over two Brillouin Zones. Additional measurements at the SIS beamline of the Swiss Light Source and the CASSIOPEE beamline of SOLEIL were used to complement this study. Fig.1 shows that the spectra are very different under LH (a) and LV (b) polarizations. In LH (also red spectra in Fig. 1c-d), a sharp peak (QP) crosses the Fermi level and a broad shoulder (HP) disperses to higher binding energies, eventually merging with a nearly non dispersive peak (NDP) centered at ~ -0.8 eV. In LV, the sharp peak and the shoulder are totally suppressed and only the NDP remains. In Fig. 1c-d, the PDH structure is emphasized at different k by subtracting the LV from LH spectra, after normalizing both spectra at -1.5 eV.

The symmetry of the orbitals probed by photoemission depends on the beam polarization, as dictated by selection rules [3]. In our experimental configuration, we expect to detect orbitals even with respect to the plane of incidence with LH (here, this is $a_{1g}$ and one $e'_g$ [4]) and odd with LV (here, the other $e'_g$). We overlay to our measurements in Fig.1a and 1b the LDA bands according to these parities. Clearly, the slope of both the QP and the HP dispersions correspond to that of the $a_{1g}$ band (see also Fig. 2a and 2b). The question arises as why there is a "break" in the $a_{1g}$ dispersion, giving rise to the QP and HP parts. Along ΓM, a large hybridization gap is predicted between $a_{1g}$ and $e'_g$, which seems to be able to produce such a situation, as proposed before [7]. We note, however, that the position expected for $e'_g$ at $k_F$ (-0.7eV) is much closer to the NDP than to the HP ($\approx$ -0.25eV, see Fig. 1d). The main problem with this explanation is that no hybridization gap is predicted along ΓK, whereas we measure almost the same PDH in the two directions. Even assuming that the hybridization gap could be larger along ΓK than in the calculation, it seems highly unlikely that it could produce a nearly identical dispersion of $a_{1g}$ near $E_F$ as that along ΓM (Fig. 2a). Therefore, we propose the alternative explanation that *the PDH is an intrinsic structure of $a_{1g}$ due to correlation effects.* The similarity between the two directions is then natural, as the band width is quite similar in the two directions.

Surprisingly, there is no clear $e'_g$ dispersions detected in these measurements. This is particularly clear in LV (Fig. 1b), where only the broad NDP is observed. This problem is analogous to the well-known absence of $e'_g$ pockets at $E_F$ in $Na_xCoO_2$ [6,7]. The $e'_g$ bands seem to be shifted away from $E_F$, which may be due to a larger crystal-field splitting between $a_{1g}$ and $e'_g$ than assumed in the calculation (this splitting changes from –10meV to 300meV, depending on the method of calculation [15]). As a result of a higher splitting, $e'_g$ bands could essentially contribute to the NDP. It is also possible that $e'_g$ bands are weak, especially at this photon energy, and somewhat hidden by some amorphous background also present in the NDP. In both cases, the subtraction of Fig.1 will be a very efficient way to reveal the true $a_{1g}$ PDH lineshape, which we now investigate.

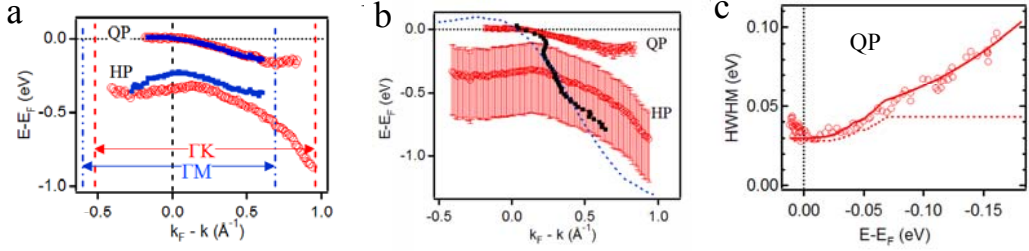

Fig. 2 : (a) Dispersion of QP and HP along ΓK (open symbols) and ΓM (closed symbols) obtained with the EDC fit of Fig. 1. (b) Comparison of the dispersion along ΓK obtained with EDC (open symbols, red) or MDC (closed symbols, black) analysis. The width of the peaks is indicated as vertical bars. The blue dotted line is the LDA dispersion for $a_{1g}$. (c) Half width at half maximum of the QP along ΓK, obtained with the EDC fit. The dotted line represents the phonon contribution and the thick line a fit adding impurity scattering (a constant term of 25meV) and electron-electron scattering (taken as $\beta\omega^2$ with $\beta=1.7eV^{-1}$).

A standard way to estimate the strength of the interactions is to calculate the effective mass $m^*$, through the renormalization of the Fermi velocity $V_F/V_{LDA}=m_{LDA}/m^*$. The dispersion can be obtained either by fitting the difference spectra at fixed k (EDC for Energy Distribution Curves) or at fixed ω (MDC for Momentum Distribution Curves) (see ref. [3] for advantages and drawbacks of the two methods). Fig. 2b compares the results of these two fits along ΓK. On the experimental side, $V_F$ is extracted through a linear fit of the dispersion in a 50meV window below $E_F$. It is quite different for the MDC ($V_F \approx 0.4eV.Å$) or EDC dispersion ($V_F \approx 0.25eV.Å$), a mismatch typical of such lineshapes [10] that we will discuss later. It is not possible to compare directly the QP and HP velocities, as the LDA dispersion is not linear over this large energy window. $V_{LDA}$ is quite different along ΓM (0.6eV.Å) and ΓK (0.85eV.Å) [4,12], mainly because of the different hybridization gaps, which may not be present in reality, as discussed before. This introduces a large incertitude on $m^*$, namely $1.5 < m^*/m_{LDA} < 3.5$, which remains compatible with rather modest interaction values, $Z \approx m^*/m_{LDA}$ being comprised between 0.3 and 0.7.

An independent estimation of Z can be obtained through direct observation of the spectral weight redistribution. The inset of Fig. 3 recalls the scheme expected for the variations of n(k), the weight integrated over ω at fixed k, in presence of correlations [3,18]. The QP weight at $k_F$ is reduced to Z, half the incoherent weight is transferred to the HP and the other half to previously unoccupied states at $k<k_F$, yielding QP/HP=2Z/(1-Z). As shown in Fig. 3, the HP is much stronger at $k_F$ than the QP and *remains strong at $k<k_F$ as typical of small Z case*. In the bottom part of Fig. 3, the area ratio between QP and HP is shown to be about 0.3 near $k_F$ for both ΓM and ΓK, yielding Z=0.15 ± 0.05 (the error bar includes estimations of fits using different HP shape near $E_F$). This value is at the very low end of the previous estimation, suggesting that the lowest $V_F$ obtained by EDC together with the highest LDA value (i.e. with no hybridization effects) are the best estimations. Remarkably, this new estimation of Z is consistent with the ratio of the QP and LDA band width ~0.2/1.2=0.17. *This gives a self-consistent view of the correlations, where 200meV directly indicates the QP energy scale.* This differentiates the PDH from a "kink" structure, where this energy would be that of a collective excitation. The spectral weight appears as a very valuable information, as it is an intimate fingerprint of the correlation strength. It is sometimes difficult to handle, because incoherent excitations may be quite diffuse, but it becomes clear here thanks to the subtraction procedure.

Regarding ARPES intensities, it is important to keep in mind that it is much more reliably defined as a function of ω than k. This is because both matrix element effects and normalization procedures [19] mainly depend on k [3,18]. Therefore, no intensity distortion is expected along one EDC (hence the ratio QP/HP is unaffected), whereas it may be strong over one MDC, especially when there is a strong intrinsic variation of n(k) [19]. In this case, we can reproduce the different $V_F$ obtained here in MDC and EDC analysis, just by assuming the extrinsic variation of I(k) sketched in Fig. 3. We suggest that the different $V_F$ reported in $Na_{0.73}CoO_2$ along ΓM and ΓK [12] or the different "kinks" reported between $Na_{0.7}CoO_2$ and BiBaCoO [13] are also likely affected by such effects, rather than by a true change of the interaction strength. The MDC analysis might therefore be quite misleading in these systems.

We now question the origin of the PDH. We first note that there are 0.3 holes left in the Co $t_{2g}$ band in BiBaCoO [5]. With a doping so close from the band insulator, residual Hubbard band should be completely suppressed [16] and it seems more natural to associate the PDH with other many-body effects. The electron-phonon coupling can be quite accurately estimated from the QP broadening as a function of binding energy [2,20] presented in Fig. 2c. The red dotted line represents the phonon contribution computed using the Eliashberg coupling function $\alpha^2F(\omega)$ obtained by first principle calculations for $Na_{0.7}CoO_2$ [21]. Clearly, it describes well the initial broadening of the peak, but is totally negligible compared to the HP width (0.3eV, see Fig. 2b). This is an experimental confirmation that the electron-phonon coupling is rather small in cobaltates. It can be described in a Debye model with a coupling constant $\lambda_D=0.2$ and a Debye frequency $\omega_D=60$meV, corresponding to $m^*=(1+\lambda_D)m_{LDA}=1.2$, i.e. Z=0.8. More generally, we have tried to reproduce the HP width, using $\lambda_D$ and $\omega_D$ as free parameters to account for the coupling with an another hypothetical bosonic mode. We can obtain reasonable fits of the width with $\lambda_D \approx 1$ and $\omega_D \approx 0.25$eV, but such values largely underestimate the HP weight, which excludes this model and underlines the importance of taking spectral weight into account to correctly describe the PDH structure. This underestimation of Z is typical of models describing kinks and confirms that a stronger type of coupling is involved here. Similar lineshapes in other systems [10,11] have been described by a polaronic model, where the HP is the envelop of many satellite excitations, arising from strong coupling between electron and one or more bosons [17]. In this particular case, the spin-orbital polaron model described by Chaloupka and Khaliullin [22] appears in good agreement with many of our observations. It would be also interesting to test other models attempting to describe the anomalous correlation effects in cobaltates [9] against the present PDH shape.

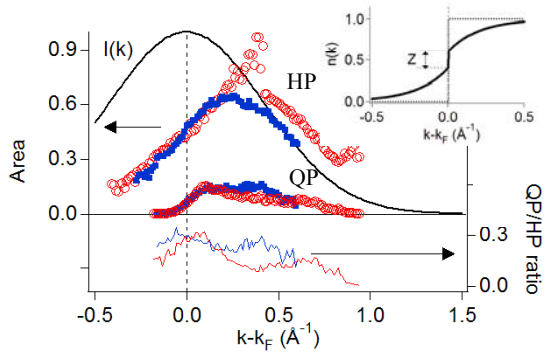

Fig. 3 : Top : Area of QP (open symbols) and HP (closed symbols) along ΓK (red circles) and ΓM (blue squares) obtained with the EDC fit of Fig. 1. Thick black line is a guess of the I(k) variation. Bottom : ratio of QP and HP area as a function of k. Inset : Sketch of the theoretical variations of n(k) for Z=0.2.

Finally, our analysis presents a few elements to characterize the interactions in BiBaCoO : the small Z value, the asymmetric lineshape of the HP, the coexistence of QP and HP over a large k-range and the transfer of significant spectral weight to the HP at $k<k_F$. One of the main puzzle in $Na_xCoO_2$ is the apparent increase of correlations at $x>0.6-0.7$ (this corresponds to 1-x holes in $t_{2g}$). This is detected by the apparition of Curie-Weiss susceptibilities and an abrupt jump of the effective mass deduced from specific heat measurements by a factor 3 to 5 [8]. Our analysis concludes that a similar, or even larger, enhancement is present at the doping equivalent to $x=0.7$ probed here, contrary to analysis based on $V_F$ values [8]. It would be interesting to revisit the evolution of the PDH at smaller $x$ with the present method to see whether it could detect a decrease of the correlations. On the high doping side ($x>0.8$), the QP peak was found tobe strongly suppressed in misfit cobaltates [5], suggesting even higher correlation effects. No similar suppression was reported so far for Na cobaltates. This may be related to the role of the potential of Na or Rock-Salt layers in building correlations through particular electronic orderings in this limit [23].

We thank J. Bobroff and D. Malterre for their critical reading of the manuscript. A. N. acknowledges the funding from the European community within the contract *Orsay Training Site in Emergent Phenomena in Condensed Matter Physics from the macro to the nanoscales*.

[1] A. Georges, G. Kotliar, W. Krauth and M.J. Rozenberg, Review of Modern Physics **68**, 13 (1996)
[2] S. LaShell, E. Jensen and T. Balasubramanian, Phys. Rev. B **61**, 2371 (2000)
[3] A. Damascelli, Z. Hussein and Z.-X. Shen, Review of Modern Physics **75**, 473 (2003).
[4] We use the band structure calculated for $Na_xCoO_2$, as it is essentially that of the $CoO_2$ plane. D.J. Singh, Phys. Rev. B **61**, 13397 (2000); K.W. Lee, J. Kand and W.E. Pickett, Phys. Rev. B **70**, 045104 (2004); H. Ishida, M.D. Johannes and A. Liebsch, Phys. Rev. Lett. **94**, 196401 (2005)
[5] V. Brouet *et al.*, Phys. Rev B 76, 100403(R) (2007)
[6] H.B. Yang *et al.*, Phys. Rev. Letters **92**, 246403 (2004); H.B. Yang *et al.*, Phys. Rev. Letters **95**, 146401 (2005)
[7] M.Z. Hasan *et al.*, Phys. Rev. Letters **92**, 246402 (2004); D. Qian *et al.*, Phys. Rev. Letters **96**, 216405 (2006); D. Qian *et al.*, Phys. Rev. Letters **97**, 186405 (2006)
[8] T.F. Schulze *et al.*, Phys. Rev. B **78**, 205101 (2008)
[9] C.A. Marianetti and G. Kotliar, Phys. Rev. Letters **98**, 176405 (2007)
[10] N. Mannella *et al.*, Nature **438**, 474 (2005)
[11] K. M. Shen *et al.*, Phys. Rev. Letters **93**, 267002 (2004)
[12] J. Geck *et al.*, Phys. Rev. Letters **99**, 046403 (2007)
[13] H.W. Ou *et al.*, Phys. Rev. Letters **102**, 026806 (2009)
[14] G. Panaccione *et al.*, Rev. Sci. Instrum. **80** 043105 (2009).
[15] C.A. Marianetti, K. Haule and O. Parcollet, Phys. Rev. Lett. 99, 246404 (2007); S. Landron and M.B. Lepetit, Phys. Rev. B **77**, 125106 (2008)
[16] A. Bourgeois, A.A. Aligia and M.J. Rozenberg, Phys. Rev. Letters **102**, 066402 (2009)
[17] The HP can be well described at all k by the same asymmetric lineshape, which we just shift. We used a Poisson distribution convoluted by a Gaussian, as in polarons [G.D. Mahan, *Many-Particle Physics* (Plenum, New York, 1981)].
[18] M. Randeria *et al.*, Phys. Rev. Letters **74**, 4951 (1995)

[19] Spectra are usually normalized as a function of k, to correct for the detector efficiency and this alone could spoil the intrinsic variation of n(k).
[20] T. Valla *et al.*, Phys. Rev. Letters **83**, 2085 (1999)
[21] J.P. Rueff et al PRB **74** 020504(R) (2006).
[22] J. Chaloupka and G. Khaliullin, Phys. Rev. Letters **99**, 256406 (2007)
[23] A. Nicolaou *et al.*, Europhysics Letters **89**, 37010 (2010)